\documentclass{elsart}
\usepackage{graphicx}
\usepackage{amssymb}

\def\be{\begin{equation}}
\def\ee{\end{equation}}
\newcommand{\dil}{{\rm dil}}
\newcommand{\pure}{{\rm pure}}

\begin{document}
\begin{frontmatter}

\title{Perturbation expansion for the diluted two-dimensional $XY$ model}

 \author[label1,label2]{O. Kapikranian,}
 \ead{akap@ph.icmp.lviv.ua}
 \author[label1]{B. Berche,}
 \ead{berche@lpm.u-nancy.fr}
 \author[label2,label3]{Yu.\ Holovatch}
 \ead{hol@icmp.lviv.ua}

 \address[label1]{Laboratoire de Physique des Mat\'eriaux, UMR CNRS 7556,\\
  Universit\'e Henri Poincar\'e, Nancy 1,\\ B.P. 239,
  F-54506  Vand\oe uvre les Nancy Cedex, France}
 \address[label2]{Institute for Condensed Matter Physics,
                  UA-79011 Lviv, Ukraine}
 \address[label3]{Institut f\"ur Theoretische Physik, Johannes Kepler
 Universit\"at Linz, A-4040 Linz, Austria}

\begin{abstract}
We study the quasi-long-range ordered phase of a 2D XY model with
quenched site-dilution using the spin-wave approximation and
expansion in the parameter which characterizes the deviation from
completely homogeneous dilution. The results, obtained by keeping
the terms up to the third order in the expansion, show good
accordance with Monte Carlo data in a wide range of dilution
concentrations far enough from the percolation threshold. We
discuss different types of expansion.
\end{abstract}

\end{frontmatter}



\noindent The XY model in two dimensions is the simplest example
of a system exhibiting ``quasi-long-range order" (QLRO), which
appears at low temperatures in a number of physical models of
great importance, e.g. magnetic films with planar anisotropy, but
also thin-film superfluids or superconductors, two-dimensional
solids, 2d-classical Coulomb gas or fluctuating surfaces and the
roughness transition \cite{Chaikin95,Nelson02}. Although no exact
solution exists for this model, most of its properties are known
from different approaches.

Destruction of long range ordering is due to the presence of
stable topological defects
(vortices)~\cite{Berezinskii71,KosterlitzThouless73,Kosterlitz74},
a situation  which strongly contrasts with usual ordering in
systems undergoing a ferromagnetic phase transition. First, the
magnetization of the XY model on an infinite 2D lattice remains
zero at any non-zero temperature~\cite{MerminWagner66}, thus it is
impossible to describe in the thermodynamic limit the
quasi-long-range ordered phase by this usual order parameter,
however the spin-spin pair correlation function gives a distinct
indication of QLRO. Its asymptotic behaviour changes from
exponential at high temperatures to power low decay at low
temperatures. This transition is referred to as the
Berezinskii-Kosterlitz-Thouless (BKT) transition and the point
where this change of behaviour occurs is the BKT temperature.

A quantity of interest which characterizes the
QLRO phase, is then the temperature dependent
exponent of the correlation function:
\begin{equation}
\eta(T) = -\lim_{|{\bf R}|\to\infty}
\frac{\ln\left<{\bf S_{\bf r}\cdot S_{r+R}}\right>}{\ln |{\bf R}|} .
\end{equation}
The spin-wave analysis of the model gives a reliable value of $\eta$ for
small enough temperatures~\cite{Wegner67}. The reliability of the harmonic
approximation for this model
is grounded by the RG analysis~\cite{Kosterlitz74}.

The case of the (classical) XY model  on a regular lattice (without defects)
\begin{equation}
H = -\frac{1}{2}\sum_{\bf r}\sum_{\bf r'}J({\bf r-r'})
\left(S^x_{\bf r}S^x_{\bf r'}+S^y_{\bf r}S^y_{\bf r'}\right),
\label{EqH_pure}
\end{equation}
has been studied intensively (when $J({\bf r-r'})$ is limited to
nearest neighbours) and its properties are well known (see e.g.
Ref.~\cite{Nelson02}). The addition of defects (site-dilution,
bond-dilution) has been considered as a trivial modification,
since  the Harris criterion \cite{Harris74} states in this case
that the universality class of the diluted model remains the same
as that of the pure one. It means that the critical exponents of
both pure and disordered models are unchanged, when evaluated at
their corresponding BKT points, e.g.
$\eta^{\dil}(T^{\dil}_{BKT})=\eta^{\pure}(T^{\pure}_{BKT})$, but
the functions $\eta^{\pure}(T)$ and $\eta^{\dil}(T)$
characterizing the low temperature phase of pure and disordered
systems are different and the exact behaviour of $\eta^{\dil}(T)$
is a question which deserves attention. For example, it is not
obvious how the impurities can interact with the vortices and
influence the QLRO. This question is addressed e.g. in
Refs.~\cite{PaulaEtAl05,WysinEtAl05}.

In a previous paper~\cite{BercheEtAl02} the influence of uncorrelated
(normally distributed) site-dilution was considered and a perturbation
expansion for the case of weak dilution was proposed. The
two-spin coupling term $J({\bf r-r'})$ in Eq.~(\ref{EqH_pure})
was replaced by
$J({\bf r-r'})c_{\bf r}c_{\bf r'}$ with
\begin{equation}
c_{\bf r} = \left\{ \begin{array}{ll}
1, & \textrm{if the site ${\bf r}$ has a spin;}\\
0, & \textrm{if the site ${\bf r}$ is empty.}
\end{array} \right.
\end{equation}
The physical quantities which characterize the system with quenched
disorder after the thermodynamical averaging must be averaged over all possible
configurations of dilution. This configurational averaging is denoted as
$\overline{(...)}$:
\begin{equation}
\overline{(...)} = \prod_{\bf r}\sum_{c_{\bf r}=0,1}[c\delta_{1-c_{\bf
r},0}+(1-c)\delta_{c_{\bf r},0}](...),\label{Eq-ConfAvg}
\end{equation}
where $c$ is the concentration of occupied sites.

Starting with the Hamiltonian in the harmonic approximation,
\begin{equation}
H = \frac{1}{4}\sum_{\bf r}\sum_{\bf r'}J({\bf r-r'})\left(\theta_{\bf r}-\theta_{\bf r'}\right)^2c_{\bf r}c_{\bf r'}\ ,
\end{equation}
and realizing the Fourier transformation of the variables:
\begin{eqnarray}
\theta_{\bf r}&=&\frac{1}{\sqrt{N}}\sum_{\bf k}e^{i{\bf
kr}}\theta_{\bf k}, \qquad \theta_{\bf
k}=\frac{1}{\sqrt{N}}\sum_{\bf r}e^{-i{\bf kr}}\theta_{\bf r},
\\
J({\bf r})&=&\frac{1}{N}\sum_{\bf q}e^{i{\bf qr}}\nu({\bf q}),
\qquad \nu({\bf q})=\sum_{\bf r}e^{-i{\bf qr}}J({\bf r}),
\end{eqnarray}
($N$ is the number of sites in the lattice, and ${\bf k}$
runs over the 1st Brillouin zone), one has
\begin{eqnarray}
\hspace{-0.5cm}H &=& J\sum_{\bf k}\gamma_{\bf k}\theta_{\bf k}\theta_{\bf -k}
+ J\sum_{\bf k}\sum_{\bf k'}(\gamma_{\bf k+k'}-\gamma_{\bf k}-\gamma_{\bf k'})
\rho({\bf k+k'})\theta_{\bf k}\theta_{\bf k'}
\\\nonumber
&+& J\sum_{\bf k}\sum_{\bf k'}\sum_{\bf q}(2-\gamma_{\bf q})\left[\rho({\bf -k-k'-q})
\rho({\bf q})-\rho({\bf -k-q})\rho({\bf -k'+q})\right]\theta_{\bf k}
\theta_{\bf k'}
\end{eqnarray}
where $\rho({\bf q}) = \frac{1}{N}\sum_{\bf r}e^{-i{\bf
qr}}(1-c_{\bf r})$, and $\gamma_{\bf k}\ \equiv\
\frac{1}{2J}\left[\nu(0) - \nu({\bf k})\right] =2-\cos k_xa-\cos
k_ya$ on the square lattice. 
The first term is the Hamiltonian of the pure system, so
one can consider $\rho$ as a parameter of perturbation of this
Hamiltonian. Note, that power of $\rho$ corresponds to the number
of sums over ${\bf k}$. A classification of the perturbation
theory series with respect to number of sums over ${\bf k}$
corresponds to the expansion in the ratio of the volume of
effective interaction to the elementary cell volume \cite{Vaks67}.
Taking this ratio to be small means that it is valid for the
short-range interacting systems, which holds for our problem.

\begin{figure} [th]
  \includegraphics[width=13.5cm]{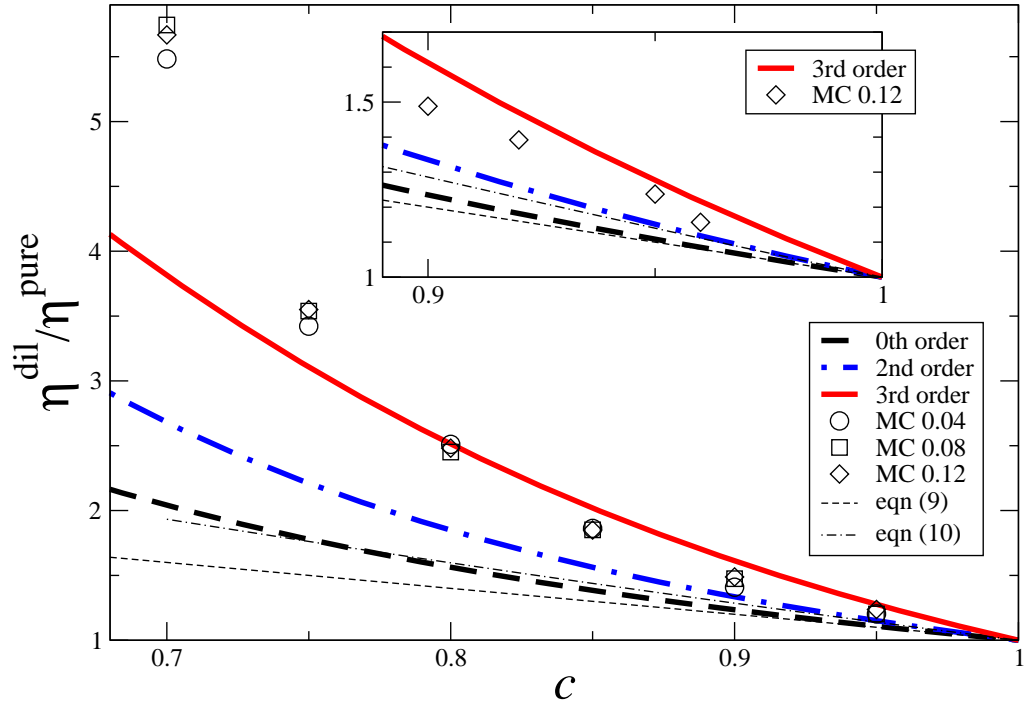}
        \caption{Comparison between the 1st (which in fact coincides with the 0th),
    2nd and 3rd-order
    expansions
     and MC simulations at
    very low temperatures (the values of $(\beta J)^{-1}$ are
    indicated in the legend).
    Expressions~(\ref{Old1st}) and (\ref{Old2nd})
    of the previous expansion are shown (thin lines)
    for comparison. Insert shows the vicinity of the pure system.}
        \label{Fig1}  \vskip -0cm
\end{figure}

The linear approximation in $\rho$-expansion presented in
Ref.~\cite{BercheEtAl02} gives the result for the exponent of the
pair correlation function:
\begin{equation}
\eta^{\dil}\ =\ \left(1+2(1-c)\right)\eta^{\pure}+O(\rho^2).\label{Old1st}
\end{equation}
Here we can report the result for this expansion up to the second order
in $\rho$:
\begin{equation}
\eta^{\dil}\ =\ \left(1+2.73(1-c)+1.27(1-c)^2\right)\eta^{\pure}
+O(\rho^3),\label{Old2nd}
\end{equation}
The figures follow from numerical estimate of sums. The 1st and
2nd order perturbation expressions fit the Monte Carlo results
only for very small con\-cen\-tra\-tions of dilution (see
Fig.~\ref{Fig1}). Unfortunately the calculation of the third order seems
to be too tedious and perhaps does not deserve so much effort,
therefore it is desirable to investigate another road.


In the present paper we propose to introduce the parameter of
expansion in a different manner in order to extend the region of
reliability of the expansion to stronger dilutions (but of course
still far enough from the percolation threshold where the whole
approach fails) and to improve convergence. We will keep the
notation $\rho$, but from now on one should understand it as the
deviation from homogeneously diluted system:
\begin{equation}
\rho({\bf q}) = \frac{1}{N}\sum_{\bf r}e^{-i{\bf qr}}(c_{\bf r}-c).
\label{eqRho}
\end{equation}
We did not make any assumption about weakness of disorder, we may
thus expect that the results of this expansion will be less
sensitive to the value of dilution $c$. Rewriting the Hamiltonian
with this new parameter one gets
\begin{eqnarray}
\hspace{-0.5cm}H &=& c^2 J\sum_{\bf k}\gamma_{\bf k}\theta_{\bf k}\theta_{\bf -k}
- cJ\sum_{\bf k}\sum_{\bf k'}(\gamma_{\bf k+k'}-\gamma_{\bf k}-\gamma_{\bf k'})
\rho({\bf k+k'})\theta_{\bf k}\theta_{\bf k'}
\nonumber \\
&+& J\sum_{\bf k}\sum_{\bf k'}\sum_{\bf q}(2-\gamma_{\bf q})\left[\rho({\bf -k-k'-q})
\rho({\bf q})-\rho({\bf -k-q})\rho({\bf -k'+q})\right]\theta_{\bf k}
\theta_{\bf k'}\nonumber\\
&\equiv& c^2 H_{\pure} + H_{\rho} + H_{\rho^2}\ ,
\label{Eq-HamDilute}
\end{eqnarray}
where the first term is the Hamiltonian of the pure system now
with a renormalized coupling.

The spin-spin pair correlation function,
\begin{equation} \label{14}
\hspace{-0.8cm}G_2(R) =
\overline{c_{\bf r}c_{\bf r+R}\left<
\cos(\theta_{\bf r+R}-\theta_{\bf r})
\right>}
= \overline{c_{\bf r}c_{\bf r+R}\frac{\left<\cos(\theta_{\bf r+R}-\theta_{\bf r})
e^{-\beta(H_{\rho}+H_{\rho^2})}\right>_*}{\left< e^{-\beta(H_{\rho}+H_{\rho^2})}\right>_*}}
\end{equation}
can be expanded in $\rho$ and then configurationally averaged. In
(\ref{14}), $\left<...\right>_*$ stands for the averaging with the
pure system Hamiltonian with renormalized coupling $c^2J$. Using
(\ref{Eq-ConfAvg}) one has the equalities: $\overline{\rho({\bf
q})}\ =\ 0$, $\overline{\rho({\bf q})\rho({\bf q'})}\ =\
c(1-c)\frac{1}{N} \delta_{{\bf q+q'},0}$ and $\overline{\rho({\bf
q})\rho({\bf q'})\rho({\bf q''})}\ =\
c(1-3c+2c^2)\frac{1}{N^2}\delta_{{\bf q+q'+q''},0}$, which,
inserted into the expansion lead after tedious calculations to the
third order expression
\begin{eqnarray}
&&\hspace{-0.8cm}\overline{c_{\bf r}c_{\bf r+R}
\left<\cos(\theta_{\bf r+R}-\theta_{\bf r})\right>}
 = c^2 \left<\cos(\theta_{\bf r+R}-\theta_{\bf r})\right>_*
\nonumber\\
&& \times\ \Bigg[\ \ 1 - \frac{1-c}{c^3}\frac{1}{\beta
JN^2}\sum_{{\bf k},{\bf k'}}\frac{(\gamma_{{\bf k}+{\bf
k'}}-\gamma_{{\bf k}}-\gamma_{{\bf k'}})^2}{\gamma^2_{\bf
k}\gamma_{\bf k'}}\sin^2\frac{\bf kR}{2}
\nonumber\\
&&
+\frac{1-3c+2c^2}{c^4}\frac{1}{\beta J}
{\textstyle
\Bigg(\frac{2}{N}\sum_{{\bf k}}
\frac{\sin^2\frac{\bf kR}{2}}{\gamma_{\bf k}}
}
\nonumber\\
&&
\textstyle
-\frac{1}{N^3}
\sum_{{\bf k},{\bf k'},{\bf k''}}\frac{(\gamma_{{\bf k}+{\bf k'}}-\gamma_{{\bf k}}-\gamma_{{\bf k'}})
(\gamma_{{\bf k'}-{\bf k''}}-\gamma_{{\bf k'}}-\gamma_{{\bf k''}})
(\gamma_{{\bf k}+{\bf k''}}-\gamma_{{\bf k}}-\gamma_{{\bf k''}})
}{\gamma^2_{\bf k}\gamma_{\bf k'}\gamma_{\bf k''}}\sin^2\frac{\bf kR}{2}
\Bigg)
\Bigg].
\label{Eq-Correl}
\end{eqnarray}
In the limit $N\rightarrow\infty$, $R\rightarrow\infty$ we have found
for the sums in (\ref{Eq-Correl}):
\begin{eqnarray}\nonumber
&&\frac{1}{N}\sum_{\bf k}\textstyle\frac{\sin^2\frac{\bf
kR}{2}}
{\gamma_{\bf k}}\ \approx\ {\rm const}\ +\ \frac{1}{2\pi}\ln\frac{R}{a},
\\
&&\frac{1}{N^2}\sum_{\bf k}\sum_{\bf k'}\textstyle{\frac{(\gamma_{\bf
k+k'}-\gamma_{\bf k}-\gamma_{\bf
k'})^2}{\gamma^2_{\bf k}\gamma_{\bf k'}}\sin^2\frac{\bf
kR}{2}}\ \approx\ {\rm const}'\ +\ \frac{0.73}{2\pi}\ln\frac{R}{a},\\\nonumber
&&\frac{1}{N^3}
\sum_{{\bf k},{\bf k'},{\bf k''}}\textstyle{\frac{(\gamma_{{\bf k}+{\bf k'}}-\gamma_{{\bf k}}-\gamma_{{\bf k'}})
(\gamma_{{\bf k'}-{\bf k''}}-\gamma_{{\bf k'}}-\gamma_{{\bf k''}})
(\gamma_{{\bf k}+{\bf k''}}-\gamma_{{\bf k}}-\gamma_{{\bf k''}})
}{\gamma^2_{\bf k}\gamma_{\bf k'}\gamma_{\bf k''}}}\sin^2\frac{\bf kR}{2}
\\\nonumber
&&\hspace{9cm}\textstyle\approx\ {\rm const}''\ +\ \frac {0.27}{2\pi}\ln\frac{R}{a}.
\end{eqnarray}
The figures $0.73$ and $0.27$ come from numerical summation. It appears that
to zeroth-order, the change of exponent comes from a renormalization
of the coupling strength, the first-order term is identically vanishing.
For small enough temperatures it is now possible to write
the pair correlation function in the power law form:
\begin{equation}
G_2(R) \approx c^2
(R/a)^{-\eta^{\dil}}\ .
\end{equation}
Reminding the correlation function exponent of the pure system in the
SW approximation, $\eta^\pure=(2\pi\beta J)^{-1}$,
we can write
\begin{equation}
\eta^{\dil} = \eta^{\pure}\left(
\frac {1}{c^2}+0.73\frac{1-c}{c^3}-0.27\frac{1-3c+2c^2}{c^4}
\right)+O(\rho^4).
\label{Eq-etabetter}
\end{equation}
The first term in the brackets, $1/c^2$, corresponds to the
zeroth order in the expansion, the first-order
term is identically vanishing as was already noted before, the
second and third terms in the brackets correspond to the second-
and third-order terms in the $\rho$-expansion respectively.

\vspace{1cm}


In order to check these expressions, simulations of 2D XY-spins
are performed using Wolff's cluster Monte Carlo
algorithm~\cite{Wolff89}. The low-temperature phase being
critical, local updates of single spins would suffer from the
critical slowing down. Implemented in the case of the XY model,
the Wolff algorithm first introduces bonds through the Ising
variables defined by the sign of the projection of the spin
variables along some random direction. Then clusters of sites are
built by a bond percolation process (here the random graph model
of the  Fortuin-Kasteleyn representation). The percolation
threshold for these bonds coincides with the Kosterlitz-Thouless
point~\cite{DukovskiMachtaChayes01}, which guarantees the
efficiency of the Wolff cluster updating scheme~\cite{Wolff89} at
$T_{\rm KT}$. In the low-temperature phase we are interested in,
this algorithm could be less efficient, but nevertheless
preferable to a local updating, since the correlation length is
diverging. Using this procedure, we discard typically $10^5$
sweeps for thermalization, and the measurements are performed on
typically $10^5$ production sweeps. Averages over disorder are
performed using typically $10^3$ samples. There is no need of a
better statistics. The boundary conditions are chosen periodic and
the critical exponent $\eta(T)$ of the correlation function is
measured indirectly through the finite-size scaling behaviour of
the magnetization \be M_T(L)\sim L^{-x_\sigma(T)},\quad
x_\sigma(T)=\frac 12\eta(T), \ee where the last scaling relations
holds in two dimensions.

In Figure~\ref{Fig1}, we compare the 0th to 3rd order expansions
(remember that the 1rd order term vanishes)
with the MC data. The agreement is quite good using the
expansion parameter~(\ref{eqRho}) which provides a clear
improvement of the previous
expansion given by expressions~(\ref{Old1st}) and (\ref{Old2nd}).
Of course the question of the next order is not settled, but now we are
on the way to counting higher orders or even summing the whole series.
Another direction of future work would be to implement the same type
of perturbation expansion within the Villain model~\cite{Villain75} and to
explore the deconfining transition of the diluted model.

\section*{Acknowledgements}

We acknowledge the CNRS-NAS exchange programme and
V. Tkachuk for interesting discussions.


\end{document}